\documentstyle[12pt,aaspp4]{article}

\begin{document}

\title{A Hybrid Cosmological Hydrodynamic/N-body Code Based on a Weighted
Essentially Non-Oscillatory Scheme}

\author{Long-Long Feng\altaffilmark{1,2,3},
       Chi-Wang Shu\altaffilmark{4}
       and Mengping Zhang\altaffilmark{5}  }

\altaffiltext{1}{Purple Mountain Observatory, Nanjing, 210008,
P.R. China. E-mail: fengll@pmo.ac.cn} \altaffiltext{2}{Center for
Astrophysics, University of Science and Technology of China,
Hefei, Anhui 230026, P.R. China} \altaffiltext{3}{National
Astronomical Observatories, Chinese Academy of Science, Chao-Yang
District, Beijing 100012, P.R. China} \altaffiltext{4}{Division of
Applied Mathematics, Brown University, Providence, RI 02912,
U.S.A. E-mail: shu@dam.brown.edu} \altaffiltext{5}{Department of
Mathematics, University of Science and Technology of China, Hefei,
Anhui 230026, P.R. China.  E-mail: mpzhang@ustc.edu.cn}

\begin{abstract}

We present a newly developed cosmological hydrodynamics code based
on weighted essentially non-oscillatory (WENO) schemes for
hyperbolic conservation laws. WENO is a higher order accurate
finite difference scheme designed for problems with piecewise
smooth solutions containing discontinuities, and has been
successfully applied for problems involving both shocks and
complicated smooth solution structures. We couple hydrodynamics
based on the WENO scheme with standard Poisson solver -
particle-mesh (PM) algorithm for evolving the self-gravitating
system. The third order low storage total variation diminishing
(TVD) Runge-Kutta scheme has been used for the time integration of
the system. To test accuracy and convergence rate of the code, we
subject it to a number of typical tests including the Sod shock
tube in multidimensions, the Sedov blast wave and formation of the
Zeldovich pancake. These tests validate the WENO hydrodynamics
with fast convergence rate and high accuracy.  We also evolve a
low density flat cosmological model ($\Lambda$CDM) to explore the
validity of the code in practical simulations.

\end{abstract}

\keywords{cosmology: theory - gravitation - hydrodynamics -
methods: numerical - shock waves}

\newpage

\section{Introduction}

Though the universe seems to be dominated by the dark sides of
both matter and energy (Turner, 2002), the observed luminous
universe has been existing in the form of baryonic matter, whose
mass density, constrained by the primordial nucleosynthesis
(Walker, et al., 1991), only occupies a small amount of the total
density. To account for the observational features revealed by the
baryonic matter, i.e., X-ray emitting gas in galaxies and clusters
(Mulchaey, 2000), intergalactic medium inferred from Ly$\alpha$
forest (Rauch, 1998), X-ray background radiation (Giacconi et al.
1962) and distorted spectrum of the cosmic background radiation
due to the Sunyaev-Zeldovich effect (Zel'dovich \& Sunyaev 1969;
Ostriker \& Vishniac 1986) etc., it would be necessary to
incorporate hydrodynamics into cosmological investigations. This
motivation has stimulated great efforts to apply a variety of gas
dynamics algorithms to cosmological simulations. For a general
review of the state-of-the-art on this topic, we refer to
Bertschinger (1998).

Due to the high non-linearity of gravitational clustering in the
universe, there are two significant features emerging in
cosmological hydrodynamic flow, which pose more challenges than
the typical hydrodynamic simulation without self-gravity. One
significant feature is the extremely supersonic motion around the
density peaks developed by gravitational instability, which leads
to strong shock discontinuities within complex smooth structures.
Another feature is the appearance of an enormous dynamic range in
space and time as well as in the related gas quantities. For
instance, the hierarchical structures in the galaxy distribution
span a wide range of length scales from a few kpc resolved by
individual galaxy to several tens of Mpc characterizing the
largest coherent scale in the universe.

A variety of numerical schemes for solving the coupled system of
collisional baryonic matter and collisionless dark matter have
been developed in the past decades. They fall into two categories,
particle methods and grid based methods.

The particle methods include variants of the smooth-particle
hydrodynamics (SPH; Gingold \& Monagham 1977, Lucy 1977) such as
those of Evrard (1988), Hernquist \& Katz (1989), Navarro \& White
(1993), Couchman, Thomas \& Pierce (Hydra, 1995), Steinmetz
(1996), Owen et al. (1998) and Springel, Yoshida \& White (Gadget,
2001). The SPH method solves the Lagrangian form of the Euler
equations, and could achieve good spatial resolutions in high
density regions, but works poorly in low density regions. It also
suffers from degraded resolution in shocked regions due to the
introduction of sizable artificial viscosity.

The grid based methods are to solve the Euler equations on
structured or unstructured grids. The early attempt was made by
Cen (1992) using a central difference scheme. It uses artificial
viscosity to handle shocks and has first-order accuracy. The
modern approaches implemented for high resolution shock capturing
are usually based on the Godunov algorithm. The two typical
examples are the total-variation diminishing (TVD) scheme (Harten
1983) and the piecewise parabolic method (PPM) (Collella \&
Woodward 1984). Both schemes start from the integral form of
conservation laws of Euler equations and compute the flux vector
based on cell averages (finite volume scheme). The TVD scheme
modifies the flux using an approximate solution of the Riemann
problem with corrections added to ensure that there are no
postshock oscillations. While in the PPM scheme, the Riemann
problem is solved accurately using a quadratic interpolation of
the cell-average densities that is constrained to minimize
postshock oscillations. In the cosmological setting, the TVD based
codes include those of Ryu et al. (1993), the moving-mesh scheme
(Pen, 1998), and the smooth Lagrangian method (Gnedin, 1995); and
the PPM based codes include those of Stone \& Norman (Zeus; 1992),
Bryan et al. (1995), Sornborger et al. (1996), Ricker, Dodelson \&
Lamb (COSMOS; 2000). The grid-based methods suffer from the
limited spatial resolution, but they work extremely well both in
low and high density regions as well as in shocks. To reach a
large dynamical range, the Godunov methods have also been
implemented with adaptive mesh refinement (RAMSES: Teyssier, 2002;
ENZO: Norman \& Bryan, 1999; O'Shea et al., 2004), which is more
adequate to explore the fine structures in the hydrodynamic
simulation.

We describe in this paper an alternative hydrodynamic solver which
discretizes the convection terms in the Euler equations by the
fifth order finite difference WENO (weighted essentially
non-oscillatory) method, first developed in Jiang \& Shu (1996),
with a low storage third order Runge-Kutta time discretization,
which was proven to be nonlinearly stable in Gottlieb \& Shu
(1998). The WENO schemes are based on the essentially
non-oscillatory (ENO) schemes first developed by Harten et al.
(1987) in the form of finite volume scheme for hyperbolic
conservative laws. The ENO scheme generalizes the total variation
diminishing (TVD) scheme of Harten (1983). The TVD schemes
typically degenerate to first-order accuracy at locations with
smooth extrema while the ENO scheme maintains high order accuracy
there even in multi-dimensions. WENO schemes further improve upon
ENO schemes in robustness and accuracy. Both ENO and WENO schemes
use the idea of adaptive stencils in the reconstruction procedure
based on the local smoothness of the numerical solution to
automatically achieve high order accuracy and non-oscillatory
property near discontinuities. For WENO schemes, this is achieved
by using a convex combination of a few candidate stencils, each
being assigned a nonlinear weight which depends on the local
smoothness of the numerical solution based on that stencil. WENO
schemes can simultaneously provide a high order resolution for the
smooth part of the solution, and a sharp, monotone shock or
contact discontinuity transition. WENO schemes are extremely
robust and stable for solutions containing strong shocks and
complex solution structures. Moreover, a significant advantage of
WENO is its ability to have high accuracy on coarser meshes and to
achieve better resolution on the largest meshes allowed by
available computer memory. We will describe the fifth order WENO
scheme employed in this paper briefly in \S 3. For more details,
we refer to Jiang \& Shu (1996) and the lecture notes by Shu
(1998, 1999).

WENO schemes have been widely used in applications. Some of the
examples include dynamical response of a stellar atmosphere to
pressure perturbations (Zanna, Velli \& Londrillo, 1998); shock
vortex interactions and other gas dynamics problems (Grasso \&
Pirozzoli, 2000a; 2000b); incompressible flow problems (Yang et
al., 1998); Hamilton-Jacobi equations (Jiang \& Peng, 2000);
magneto-hydrodynamics (Jiang \& Wu, 1999); underwater blast-wave
focusing (Liang \& Chen, 1999); the composite schemes and shallow
water equations (Liska \& Wendroff, 1998, 1999); real gas
computations (Montarnal \& Shu, 1999), wave propagation using
Fey's method of transport (Noelle, 2000); etc.

In the context of cosmological applications, we have developed a
hybrid N-body/hydrodynamical code that incorporates a Lagrangian
particle-mesh algorithm to evolve the collisionless matter with
the fifth order WENO scheme to solve the equations of gas
dynamics. This paper is to detail this code and assess its
accuracy using some numerical tests. We proceed as follows. In \S
2, we present the basic cosmological hydrodynamic equation for the
baryon-CDM coupling system. \S 3 gives a brief discussion of the
numerical scheme for solving the hydrodynamic equations,
especially about the implementation of the finite difference fifth
order WENO scheme and the TVD time discretization. In \S 4, we
validate the code using a few challenging numerical tests.
Concluding remarks are drawn in \S 5.

\section{The Basic Equations}

The hydrodynamic equations for baryons in the expanding universe,
without any viscous and thermal conductivity terms, can be written
in the following compact form,
\begin{equation}  \label{hydro}
U_t+f(U)_X+g(U)_Y+h(U)_Z=F(t,U)
\end{equation}
where $U$ and the fluxes $f(U)$, $g(U)$ and $h(U)$ are
five-component column vectors,
\begin{equation}
\left(
\begin{array}{c}
\rho \\
\rho u \\
\rho v \\
\rho w \\
E%
\end{array}
\right) , \qquad \left(
\begin{array}{c}
\rho u \\
\rho u^2+P \\
\rho uv \\
\rho uw \\
u(E+P)%
\end{array}
\right) , \qquad \left(
\begin{array}{c}
\rho v \\
\rho uv \\
\rho v^2+P \\
\rho vw \\
v(E+P)%
\end{array}
\right) , \qquad \left(
\begin{array}{c}
\rho w \\
\rho uw \\
\rho vw \\
\rho w^2+P \\
w(E+P)%
\end{array}
\right) .
\end{equation}
Here t is the cosmic time, $\mathbf{R}=(X,Y,Z)$ is the proper
coordinates,
which is related to comoving coordinates $\mathbf{r}=(x,y,z)$ via $\mathbf{R}%
=a(t)\mathbf{r}$; the subscripts ($X,Y,Z$) in equation
(\ref{hydro}) denote spatial derivatives, e.g.
$(\cdot)_{X}=\partial(\cdot)/\partial X$; $a(t)$
is the expansion scale factor, $\rho$ is the comoving density, $\mathbf{V}%
=(u,v,w)$ is the proper peculiar velocity, E is the total energy
including both kinetic and internal energies, P is comoving
pressure, which is related to the total energy E by
\begin{equation}
E=\frac{P}{\gamma-1}+\frac{1}{2}\rho(u^2+v^2+w^2)
\end{equation}
where we assume an ideal gas equation of state, $P=(\gamma-1)e$,
where $e$ is the total internal energy and $\gamma$ is the ratio
of the specific heats of the baryon; for a monatomic gas,
$\gamma=5/3$. The left hand side of equation (\ref{hydro}) is
written in the conservative form for mass, momentum and energy,
the ``force'' source term on the right hand side includes the
contributions from the expansion of the universe and the
gravitation:
\begin{equation}
\left(
\begin{array}{c}
0 \\
-\frac{\dot a}{a}\rho\mathbf{V}+\rho\mathbf{G} \\
-2\frac{\dot a}{a}E+\rho\mathbf{V}\cdot\mathbf{G}-\Lambda_{net}%
\end{array}
\right) ,
\end{equation}
where $\Lambda_{net}$ represents the net energy loss due to the
radiative
heating-cooling of the baryonic gas, and $\mathbf{G}=-\mathbf{\nabla}_{%
\mathbf{R}}\Phi$ is the peculiar acceleration in the gravitational
field produced by both the dark matter and the baryonic matter.

The motions of the collisionless dark matter in comoving
coordinates are governed by a set of Newtonian equations,
\begin{eqnarray}  \label{DMmotion}
\frac{d\mathbf{r}_{DM}}{dt} & = & \frac{1}{a}\mathbf{v}_{DM}  \nonumber \\
\frac{d\mathbf{v}_{DM}}{dt} & = & -\frac{\dot
a}{a}\mathbf{v}_{DM}+\mathbf{G}
\end{eqnarray}
where $\mathbf{r}_{DM}$ and $\mathbf{v}_{DM}$ are the comoving
coordinates and the proper peculiar velocity respectively, and the
subscript $DM$ refers to the dark matter. The peculiar
gravitational potential obeys the Poisson equation,
\begin{equation}
\nabla^2 \Phi(\mathbf{x},t)=4\pi
G[\rho_{tot}(\mathbf{x},t)-\rho_0(t)]/a
\end{equation}
in which $G$ is the gravitational constant,
$\rho_{tot}=\rho_{b}+\rho_{DM}$ is a sum of the comoving baryon
and dark matter density, and $\rho_0(t)$ is the uniform background
density at time $t$.

\section{Numerical Techniques}

\subsection{Hydrodynamic Solver: Finite Difference WENO Schemes}

\subsubsection{Approximating the derivatives}

The fifth order WENO finite difference spatial discretization to a
conservation law such as
\begin{equation}  \label{1.1}
u_t + f(u)_x + g(u)_y + h(u)_z = 0
\end{equation}
approximates the derivatives, for example $f(u)_x$, by a
conservative difference
\[
f(u)_x |_{x=x_j} \approx \frac{1}{\Delta x} \left( \hat{f}_{j+1/2} - \hat{f}%
_{j-1/2} \right)
\]
along the $x$ line, with $y$ and $z$ fixed, where
$\hat{f}_{j+1/2}$ is the numerical flux. $g(u)_y$ and $h(u)_z$ are
approximated in the same way. Hence finite difference methods have
the same format for one and several space dimensions, which is a
major advantage. For the simplest case of a scalar equation
(\ref{1.1}) and if $f^{\prime}(u) \geq 0$, the fifth order finite
difference WENO scheme has the flux given by
\[
\hat{f}_{j+1/2} = w_1 \hat{f}_{j+1/2}^{(1)} + w_2
\hat{f}_{j+1/2}^{(2)} + w_3 \hat{f}_{j+1/2}^{(3)}
\]
where $\hat{f}_{j+1/2}^{(i)}$ are three third order accurate
fluxes on three different stencils given by
\begin{eqnarray*}
\hat{f}_{j+1/2}^{(1)} & = & \frac{1}{3} f(u_{j-2}) - \frac{7}{6}
f(u_{j-1})
+ \frac{11}{6} f(u_{j}), \\
\hat{f}_{j+1/2}^{(2)} & = & -\frac{1}{6} f(u_{j-1}) + \frac{5}{6}
f(u_{j}) +
\frac{1}{3} f(u_{j+1}), \\
\hat{f}_{j+1/2}^{(3)} & = & \frac{1}{3} f(u_{j}) + \frac{5}{6}
f(u_{j+1}) - \frac{1}{6} f(u_{j+2}).
\end{eqnarray*}
Notice that the combined stencil for the flux $\hat{f}_{j+1/2}$ is
biased to the left, which is upwinding for the positive wind
direction due to the assumption $f^{\prime}(u) \geq 0$. The key
ingredient for the success of WENO scheme relies on the design of
the nonlinear weights $w_i$, which are given by
\[
w_i = \frac {\tilde{w}_i}{\sum_{k=1}^3 \tilde{w}_k},\qquad
\tilde{w}_k = \frac {\gamma_k}{(\varepsilon + \beta_k)^2} ,
\]
where the linear weights $\gamma_k$ are chosen to yield fifth
order accuracy when combining three third order accurate fluxes,
and are given by
\[
\gamma_1=\frac{1}{10}, \qquad \gamma_2=\frac{3}{5}, \qquad \gamma_3=\frac{3}{%
10} ;
\]
the smoothness indicators $\beta_k$ are given by
\begin{eqnarray*}
\beta_1 & = & \frac{13}{12} \left( f(u_{j-2}) - 2 f(u_{j-1}) +
f(u_{j}) \right)^2 + \frac{1}{4} \left( f(u_{j-2}) - 4 f(u_{j-1})
+ 3 f(u_{j})
\right)^2 \\
\beta_2 & = & \frac{13}{12} \left( f(u_{j-1}) - 2 f(u_{j}) +
f(u_{j+1})
\right)^2 + \frac{1}{4} \left( f(u_{j-1}) - f(u_{j+1}) \right)^2 \\
\beta_3 & = & \frac{13}{12} \left( f(u_{j}) - 2 f(u_{j+1}) +
f(u_{j+2}) \right)^2 + \frac{1}{4} \left( 3 f(u_{j}) - 4
f(u_{j+1}) + f(u_{j+2}) \right)^2 ,
\end{eqnarray*}
and they measure how smooth the approximation based on a specific
stencil is in the target cell. Finally, $\varepsilon$ is a
parameter to avoid the denominator to become 0 and is usually
taken as $\varepsilon = 10^{-6}$ in the computation. The choice of
$\varepsilon$ does not affect accuracy, the errors can go down to
machine zero with mesh refinement while $\varepsilon = 10^{-6}$ is
kept fixed.

This finishes the description of the fifth order finite difference
WENO scheme in Jiang \& Shu (1996) in the simplest case. As we can
see, the algorithm is actually quite simple and the user does not
need to tune any parameters in the scheme.

\subsubsection{Properties of the WENO scheme}

We briefly summarize the properties of this WENO finite difference
scheme. For details of proofs and numerical verifications, see
Jiang \& Shu (1996) and the lecture notes of Shu (1998, 1999).

\begin{enumerate}
\item The scheme is proven to be uniformly fifth order accurate
including at smooth extrema, and this is verified numerically.

\item Near discontinuities the scheme produces sharp and
non-oscillatory discontinuity transition.

\item The approximation is self-similar. That is, when fully
discretized with the Runge-Kutta methods in next section, the
scheme is invariant when the spatial and time variables are scaled
by the same factor. This is a major advantage for approximating
conservation laws which are invariant under such scaling.
\end{enumerate}

\subsubsection{Generalization to more complex situations}

We then indicate how the scheme is generalized in a more complex
situation, eventually to 3D systems such as the Euler equations:

\begin{enumerate}
\item For scalar equations without the property $f^{\prime}(u)
\geq 0$, one uses a flux splitting
\[
f(u)= f^+(u) + f^-(u), \qquad \frac{d f^+(u)}{du} \geq 0, \,\,\,
\frac{d f^-(u)}{du} \leq 0,
\]
and apply the above procedure to $f^+(u)$, and a mirror image
(with respect to $j+1/2$) procedure to $f^-(u)$. The only
requirement for the splitting is that $f^\pm(u)$ should be smooth
functions of $u$. In this paper we use the simple Lax-Friedrichs
flux splitting
\[
f^\pm(u) = \frac{1}{2} ( f(u) \pm \alpha u), \qquad \alpha =
max_{u} |f^{\prime}(u)|
\]
where the maximum is taken over the relevant range of $u$. This
simple Lax-Friedrichs flux splitting is quite diffusive when
applied to first and second order discretizations, but for the
fifth order WENO discretization we adopt, it has very small
numerical viscosity.

\item For systems of hyperbolic conservation laws, the nonlinear
part of the WENO procedure (i.e. the determination of the
smoothness indicators $\beta_k$ and hence the nonlinear weights
$w_i$) is carried out in local
characteristic fields. Thus one would first find an average $u_{j+1/2}$ of $%
u_j$ and $u_{j+1}$ (we use the Roe average, Roe (1978), which
exists for many physical systems including the Euler equations),
and compute the left and right eigenvectors of the Jacobian
$f^{\prime}(u_{j+1/2})$ and put them into the rows of a matrix
$R^{-1}_{j+1/2}$ and the columns of another matrix $R_{j+1/2}$,
respectively, such that $R^{-1}_{j+1/2} \, f^{\prime}(u_{j+1/2})
\, R_{j+1/2} = \Lambda_{j+1/2}$ where $\Lambda_{j+1/2}$
is a diagonal matrix containing the real eigenvalues of $f^{%
\prime}(u_{j+1/2})$. One then transforms all the quantities needed
for evaluating the numerical flux $\hat{f}_{j+1/2}$ to the local
characteristic fields by left multiplying them with
$R^{-1}_{j+1/2}$, and then computes the numerical fluxes by the
scalar procedure in each characteristic field. Finally, the flux
in the original physical space is obtained by left multiplying the
numerical flux obtained in the local characteristic fields with
$R_{j+1/2}$.

\item If one has a non-uniform but smooth mesh, for example
$x=x(\xi)$ where $\xi_j$ is uniform and $x(\xi)$ is a smooth
function of $\xi$, then one could use the chain rule $f(u)_x =
f(u)_\xi / x^{\prime}(\xi)$ and simply use the procedure above for
uniform meshes to approximate $f(u)_\xi$. Using this, one could
use finite difference WENO schemes on smooth curvilinear
coordinates in any space dimension.

\item WENO finite difference schemes are available for all odd
orders, see Liu, Osher \& Chan (1994) and Balsara and Shu (2000)
for the formulae of the third order and seventh through eleventh
order WENO schemes.
\end{enumerate}

\subsection{Time Discretizations}

The finite difference WENO scheme we use in this paper is
formulated first as method of lines, namely discretized in the
spatial variables only. It is still necessary for us to discretize
the time variable. Often it is easier to prove stability (e.g. for
TVD schemes) when the time variable is discretized by the first
order accurate forward Euler, however time accuracy is as
important as spatial accuracy, hence we would like to have higher
order accuracy in time while maintaining the stability properties
of the forward Euler time stepping. We use a class of high order
nonlinearly stable Runge-Kutta time discretizations. A distinctive
feature of this class of time discretizations is that they are
convex combinations of first order forward Euler steps, hence they
maintain strong stability properties in any semi-norm (total
variation norm, maximum norm, entropy condition, etc.) of the
forward Euler step, with a time step restriction proportional to
that for the forward Euler step to be stable, this proportion
coefficient being termed CFL (Courant-Friedrichs-Levy, referring
to stability restrictions on the time step) coefficient of the
high order Runge-Kutta method. Thus one only needs to prove
nonlinear stability for the first order forward Euler step, which
is relatively easy in many situations (e.g. TVD schemes), and one
automatically obtains the same strong stability property for the
higher order time discretizations in this class. These methods
were first developed in Shu \& Osher (1988) and Shu (1988), and
later generalized in Gottlieb \& Shu (1998) and Gottlieb, Shu \&
Tadmor (2001) . The most popular scheme in this class is the
following third order Runge-Kutta method for solving
\[
u_t = L(u,t)
\]
where $L(u,t)$ is a spatial discretization operator (it does not
need to be, and often is not, linear):
\begin{eqnarray}  \label{s1}
u^{(1)} & = & u^n + \Delta t L(u^n, t^n)  \nonumber \\
u^{(2)} & = & \frac{3}{4} u^n + \frac{1}{4} u^{(1)} + \frac{1}{4}
\Delta t
L(u^{(1)}, t^n+\Delta t) \\
u^{n+1} & = & \frac{1}{3} u^n + \frac{2}{3} u^{(2)} + \frac{2}{3}
\Delta t L(u^{(2)}, t^n+ \frac{1}{2} \Delta t ) ,  \nonumber
\end{eqnarray}
which is nonlinearly stable with a CFL coefficient 1. However, for
our purpose of 3D calculations, storage is a paramount
consideration. We thus use a third order low storage nonlinearly
stable Runge-Kutta method, which was proven to be nonlinearly
stable with a CFL coefficient 0.32 (Gottlieb \& Shu, 1998).
Although the time step of this low storage method must be smaller
for stability analysis, in practice the time step can be taken
larger (for example with a CFL coefficient 0.6 used in this paper)
without observing any instability. In appendix A, the algorithm of
the third-order low storage Runge-Kutta method is given. This
method is to be applied to the numerical tests presented in the
following section (\S 4).

\subsection{Resolving the High Mach Number Problem}

In cosmological hydrodynamic simulations, one main challenge is to
track precisely the thermodynamic evolution in supersonic flows
around the density peaks due to gravitational collapse. The
supersonic flow could have a high Mach number, as large as
$M\sim100$, at which the ratio of the internal
thermal energy $E_{th}$ to the kinetic energy $E_k$ is as small as $%
M^{-2}\sim10^{-4}$. In an Eulerian numerical scheme for
hydrodynamics, the thermal energy is obtained by subtracting the
kinetic energy from the total energy $E_{th}=E-E_{k}$. This
calculation leads to a significant error if the thermal energy is
negligibly small comparing with the kinetic energy. Even though
there is an improvement of the quality when WENO scheme is used,
due to its high order accuracy near shock fronts, the problem
still remains. This is what is referred to in the literature as
the high Mach number problem.

To tackle the high Mach number flow that frequently appears in
cosmological hydrodynamic simulations, the current common practice
is to solve the thermal energy accurately using a complementary
equation in the unshocked region, either a modified entropy
equation (Ryu et al., 1993) or the internal energy equation (Bryan
et al., 1995). In this paper, we combine these two approaches.
That is, we take the dual energy approach of Bryan et al., but
instead of solving the internal energy equation, we follow Ryu et
al. (1993) to solve the modified entropy equation. Without taking
account of the energy loss across shocks, the conservative form of
the modified entropy is,
\begin{equation}  \label{entropy}
\frac{\partial S}{\partial t}+\frac{1}{a}\mathbf{\nabla}\cdot(S \mathbf{V}%
)=-2\frac{\dot a}{a}S
\end{equation}
where $S$ is the modified entropy defined by $S \equiv
p/\rho^{\gamma-1}$. It is noted that the entropy equation is only
valid in unshocked regions, and can be solved numerically by the
standard WENO finite difference scheme. With the entropy equation
(\ref{entropy}), the thermal energy is updated from the results of
either total energy inside shocks or modified entropy outside
shocks according to an ad hoc criterion, which operates on each
cell using
\begin{eqnarray}
p=\left\{
\begin{array}{l@{\quad}r}
(\gamma-1)(E-\frac{1}{2}\rho v^2), & (E-\frac{1}{2}\rho v^2)/E \geq \eta \\
S\rho^{\gamma-1}, & (E-\frac{1}{2}\rho v^2)/E<\eta.%
\end{array}
\right.
\end{eqnarray}
where $\eta$ is a free parameter. We take $\eta=10^{-3}$ in our
calculations in order to have no noticeable dynamical effect on
the system. To incorporate the pressure obtained from the modified
entropy equation into the total energy equation, we reset the
total energy $E$ or the entropy $S$ at each time loop. Namely,
according to the criterion equation (10), if the pressure is
determined by the total energy equation, we update the entropy by
$S=p/\rho^{\gamma-1}$; and if the pressure is given by the entropy
equation, we reset the total energy by $E=p/(\gamma-1)+\rho
v^2/2$. These procedures enable us to track both the thermal
energy and total energy correctly in the shocked and unshocked
regions.

\subsection{Implementation}

In practical cosmological simulations, the code proceeds according
to the following stages:

\begin{enumerate}
\item Under the Gaussian assumption of the primordial density
fluctuations, we initialize a simulation using the Zeldovich
approximation to set up a distribution of CDM particles. The
baryonic density and velocity fields are then given as in Cen
(1992);

\item The WENO scheme is applied to compute the advection fluxes
for the hydrodynamic variables as described in \S 3.1;

\item The gravitational field is solved by the standard
particle-mesh N-body technique (See Hockney \& Eastwood, 1988;
Efstathiou et al. 1985). Namely, for the dark matter particles,
the density is assigned to the grid with a cloud-in-cell (CIC)
method and then subjected to a Fast Fourier Transform (FFT) to
generate the discretized density field; the gravitational
potential to the Poisson equation is then obtained by a
convolution technique, in which we make use of the optimized Green
function appropriate to the seven-point finite difference
approximation to the Laplacian;

\item The positions and velocities of CDM particles as well as the
hydrodynamic variables are updated with the third order low
storage Runge-Kutta method, which ensures third order accuracy in
the time integration of the system.
\end{enumerate}

The time step is chosen by the minimum value among three time
scales. The
first is from the Courant condition given by 
\begin{equation}
\Delta t \le \frac{ CFL \times a(t) \Delta
x}{\hbox{max}(|u_x|+c_s, |u_y|+c_s, |u_z|+c_s)}
\end{equation}
where $\Delta x$ is the cell size, $c_s$ is the local sound speed, $u_x$, $%
u_y$ and $u_z$ are the local fluid velocities and $CFL$ is the
Courant number for the stability of time discretization. The
analysis for nonlinear stability allows the Courant number to be
up to 1 for the regular third
order nonlinearly stable Runge-Kutta time discretization given by equation (%
\ref{s1}), and up to 0.32 for the low storage third order
nonlinearly stable Runge-Kutta time discretization given by
equations (\ref{s2})-(\ref{s3}), that we use in this paper.
Typically, we take $CFL=0.6$ in our computation and observe stable
results. The second constraint is imposed by cosmic expansion
which requires that $\Delta a /a <0.02$ within a single time step.
This constraint comes from the requirement that a particle moves
no more than a fixed fraction of the cell size in one time step.
In cosmological simulations, the time step is always controlled by
the cosmological expansion at the early stage of evolution, but
most of the CPU time is spent in Courant time steps at the later
nonlinear clustering regimes.

Our hybrid cosmological hydrodynamic/N-body code has been written
in Fortran 90. Compiling on a DELL precision 530 workstation with
one Intel(R) Xeon(TM) Processor 2.8GHz/533MHz, it runs at the
speed of $\sim 1.9\times 10^4$ zones per second without the
N-body/gravity solver and $\sim 1.6\times 10^4$ zones with the
N-body/gravity solver. For the benchmark of the WENO code with the
same implementation as ours but without gravity, performed on an
IBM SP parallel computer, we refer to Tables (1) and (2) of Shi,
Zhang \& Shu (2003) for details.

\section{Numerical Tests}

The WENO scheme in application to both compressible and
incompressible gas hydrodynamics has been subjected to a variety
of numerical tests, e.g., shock tube problem, Double Mach
reflection, 2-dimensional shock vortex interactions, etc. All of
these tests work very well, especially for the situation when both
shocks and complicated smooth flow features co-exist,
demonstrating the advantages of high order schemes. For these test
results of WENO schemes, see, e.g. Shu (2003) and the references
therein. In this section, we are going to run the following tests:
(1) the Sod shock tube tests in one-, two- and three-dimensions;
(2) the Sedov spherical blast wave in 3-dimensions; (3) the
Zeldovich pancake which characterizes the structure formation in
the universe by the single-mode analysis; and (4) finally, we
demonstrate the code by simulating the adiabatic evolution of the
universe in a $\Lambda$CDM model.

\subsection{Shock Tube Test}

The Sod shock tube problem (Sod, 1978) has been widely used to
test the ability of hydrodynamic codes for shock capturing. Under
a specifically chosen initial condition, it could produce all of
three types of fluid discontinuity: shock, contact and
rarefaction. The Sod problem is set as a straight tube of gas
divided by a membrane into two chambers. The initial state of the
gas are specified by uniform density and pressure on both chambers
respectively. On the left chamber, we set $\rho_L=1.5$, $p_L=1.0$,
and on the right, $\rho_R=1.0$, $p_R=0.2$. The gas is assumed to
be at rest everywhere initially. The polytropic index is
$\gamma=1.4$.

The Sod shock tube is actually a 1-dimensional problem. To find
how well the shock structure is resolved in high dimensions, we
perform the test in one-, two- and three-dimensional cases. In
1-dimension, the shock propagates along the line of the x-axis.
For two- and three-dimensional cases, the shock propagates along
the main diagonal of the calculation region, i.e., along the line
(0,0) to (1,1) in the square and along the line (0,0,0) to (1,1,1)
in the cube. Fig. 1 compares the numerical results at $t=0.195$
with the analytical solution in one-, two- and three-dimensions
respectively, in which 64 cells in each direction have been used.
Comparable to the calculations done with PPM or TVD schemes (e.g.
Ryu et al. 1993, Pen 1998, Ricker et al., 2000), the shock and
contact discontinuity can be resolved within two to three cells in
the multidimensional calculations, and moreover, each quantity
gets some improvements with the increase of spatial dimensions.

\subsection{Spherical Sedov-Taylor Blast Wave}

Another challenging test for the 3-dimensional hydrodynamic code
is the Sedov blast wave (Sedov, 1993). We initialize the
simulation by setting up a point-like energy release in a
homogeneous medium of density and negligible pressure. This
explosion will develop a spherical blast wave that sweeps material
around as it propagates outward along the radial direction. The
derivation of the full analytical solutions can be found in Landau
\& Lifshitz (1987). It has been currently used for modeling the
supernova explosion.

The shock front propagates according to
\begin{equation}
r_s(t)=\xi_0 \left( \frac{E_0t^2}{\rho_1} \right)^{1/5}
\end{equation}
where $\xi_0=1.15$ for an ideal gas with a polytropic index
$\gamma=5/3$. The velocity of the shock is given by $v_s=\partial
r_s(t)/\partial t$. Behind the shock, the density, momentum and
pressure are given by
\begin{eqnarray}
\rho_2 &=& \frac{\gamma+1}{\gamma-1}\rho_1 \\
(\rho v)_2 &=& \frac{2}{\gamma-1} \rho_1 v_{s} \\
P_2 &=& \frac{2}{\gamma+1}\rho_1v_s^2 .
\end{eqnarray}

We apply the 3-dimensional WENO scheme to run the Sedov-Taylor
blast test. The simulation is performed in a cubic box with a
$256^3$ grid, and initialized by setting up a uniform density
$\rho_1=1$ and negligible pressure with a very small value
$p_0=10^{-5}$ to match a numerical approximation to zero. A
point-like energy $E_0=10^5$ is injected at the center of the box,
and the medium is at rest initially. The challenging nature of the
spherical Sedov-Taylor blast wave stems from the fact that a
Cartesian grid is used. To minimize the anisotropic effects due to
the Cartesian coordinates, we convolve the initial condition with
a spherical Gaussian filter with a window radius of 1.5 grids.

The full three-dimensional numerical solutions for density,
momentum and pressure are displayed in Fig. 2 by projecting onto
the radial coordinate. As can be seen in Fig. 2, the numerical
solution captures the spatial profile of the shock well, although
there is still some scattering around the analytical solution.
Obviously, the scattering originates from the geometric effect of
the projection from the Cartesian grid onto the spherical
coordinate. Such geometric anisotropy also leads to the shock
front being not fully resolved within one cell as described by the
analytical solution. Accordingly, in the widened shock front of
the numerical solution, both density and pressure have been
underestimated in comparison with their predicted maximum values.
Fig. 3 presents the density distributions in a slice across the
explosion point. The anisotropy could be clearly seen in this
figure.

\subsection{Zeldovich Pancake}

The Zeldovich pancake problem (Zeldovich, 1970) provides a
stringent numerical test for cosmological hydrodynamic codes. It
involves the basic physics underlying in cosmological simulation,
namely, hydrodynamics, self-gravity, cosmic expansion, and strong
shock formed in smooth structure with high Mach numbers. In the
one-dimensional case, the problem can be formulated by placing a
sinusoidal perturbation along the axis and tracking its evolution.
In the linear or quasi-linear regime, there exists an exact
solution in the Lagrangian coordinate if the pressure is
neglected. For a flat cosmology, the solution can be written in
the following forms
\begin{eqnarray}  \label{za}
\rho(x_l)&=&\rho_0\Big[1-\frac{1+z_c}{1+z}\cos(kx_l)\Bigr]^{-1} \\
v(x_l)&=&-H_0\frac{1+z_c}{(1+z)^{1/2}}\frac{\sin(kx_l)}{k}
\end{eqnarray}
where $z_c$ is a redshift at which the gravitational collapse
results in the formation of caustics, i.e., Zeldovich pancake.
$H_0$ is the Huuble constant, $H_0=100h km/s/Mpc$, and
$k=2\pi/\lambda$ specifies the comoving perturbation wavelength.
$x_l$ is the Lagrangian coordinate which is related to the
Eulerian position $x_e$ by
\begin{equation}
x_l-\frac{1+z_c}{1+z}\frac{\sin(kx_l)}{k}=x_e .
\end{equation}

For the numerical model, we adopt the same parameters as those
used in Bryan
et al. (1995), which are given by $z_c=1$, $\Omega=1$ , $h=0.5$, $%
\lambda=64h^{-1}Mpc$. The simulation is performed from the initial redshift $%
z_i=100$ for the purely baryonic gas with a uniform temperature
distribution $T_i=100K$.

It is noted that the analytical solution given by equation
(\ref{za}) holds until the redshift $z=z_c$ of caustic formation.
In Fig. 4, we compare the numerical solution using a $256$ grid to
the Zeldovich pancake with the analytical solution at z=10 and
z=1.05. We can clearly see an excellent agreement between the
numerical and analytical solutions. To access the accuracy to
which our WENO/PM code is able to reach, we run the code with a
fixed perturbation wavelength but a varying number of zones $N$.
Using the exact solution (16)-(18), we define the $L^1$ error
norms for the density as
\begin{equation}
\Delta \rho = \frac{1}{N} \sum_{i=1}^{N} |\frac{\rho_i-\rho_{zel,i}}{%
\rho_{zel,i}}|
\end{equation}
where $\rho_i$ is the numerical solution on the grid, and
$\rho_{zel,i}$ is the Zeldovich solution given by equation (16).
The $L^1$ error norm for the velocity fields are defined
similarly. The results at different redshifts ranging from the
linear regime $z=20$ to the highly nonlinear collapsing phase
$z=1.05$ are displayed in Fig. 5. Usually, the $L^1$ error should
be scaled with $N$ according to a power law $\sim N^{-r}$ or with
spatial resolutions as ${\Delta x}^r$, where $r$ is defined as the
convergence rate. For the density field, the $L^1$ error declines
rapidly with $r\simeq 1.8$ at $z=20$. With decreasing redshifts,
the convergence rate slows down, e.g., the error varies
approximately as $N^{-1}$ at $z=1.05$, namely, roughly a linear
convergence law with the spatial resolution. The velocity at
$z=20$ converges somewhat faster than the density with convergence
rate $r\simeq 1.9 $, but somewhat slower at $z=1.05$ with $r\simeq
0.9$.

The non-linear evolution subsequent to the caustics formation is
more difficult to track numerically than that in the linear phase.
The formation of the caustics is due to the head-on collision of
two cold bulk flows and gravitational collapsing, which result in
a strong shock and large gradients in the involved physical
fields. Moreover, a large range of variation in the temperature
distribution is also difficult to capture numerically. In Fig. 6
we plot the solution for the density, velocity and temperature
distribution at $z=0$ obtained from a 256 zone run. It should be
noted that, in the unshocked region, the temperature is solved
from the entropy equation and remains with a uniform temperature
1K, which is the artificial minimum temperature.

To determine how well the shock is resolved by the WENO scheme at
different resolutions, we also make runs with 32, 64, 128 zones
and compare with a high resolution run with 512 zones. Unlike the
similar comparison done by Bryan using the PPM scheme, we have not
degraded the solution at the high resolution to appropriate
scales. The results are shown in Fig. 7. Clearly, the shock
structure is well resolved by approximately equal number of zones
for the three low resolution solutions, although the width of the
shock is widened with reduced resolution correspondingly.
Moreover, we see that the solution with 128 zones has already
converged to that with 512 zones, which is likely to be the real
physical solution to the problem. This demonstrates the rapid
convergence rate of the high order WENO scheme even in the highly
nonlinear phase of the caustics.

\subsection{A Cosmological Application: the $\Lambda$CDM Model}

We run the hybrid hydrodynamic/N-body WENO/PM code to track the
cosmic evolution of the coupled system of both dark matter and
baryonic matter in a flat low density CDM model ($\Lambda$CDM),
which is specified by the density parameter $\Omega_m=0.3$,
cosmological constant $\Omega_{\Lambda}=0.7$, Hubble constant
$h=0.7$, and the mass fluctuation within a sphere of radius
8h$^{-1}$Mpc, $\sigma_8=0.9$. The baryon fraction is fixed using
the constraint from primordial nucleosynthesis as
$\Omega_b=0.0125h^{-2}$ (Walker et al., 1991). The initial
condition has been generated by the Gaussian random field with the
linear CDM power spectrum taken from the fitting formulae
presented by Eisenstein \& Hu (1998). The simulation is performed
in a periodic box of side length of 25h$^{-1}$Mpc with a 192$^3$
grid and an equal number of dark matter particles. The universe is
evolved from $z=49$ to $z=0$. The initial temperature is set to
$T=10^4K$ and the polytropic index takes the value $\gamma=5/3$.
For comparison, two sets of simulation have been performed by
using the WENO-E and WENO-S schemes, which update the energy and
entropy respectively outside the shocked regions, see section 3.3.

Fig. 8 plots density contours for the baryonic matter (lower
panel) and the
cold dark matter (upper panel) in a slice with 2-cell thickness of 0.26h$%
^{-1}$Mpc at $z=1.5$. In Fig. 9, we compare the cell temperature
contours drawn from the simulations using the WENO-E (lower panel)
and WENO-S (upper panel) codes. Obviously, the two simulations
coincide in the high temperature $(>10^4K)$ regions, but in the
low temperature regions, the WENO-E simulation gives less
fractions of the volume than those from WENO-S. This phenomenon is
due to the artificial numerical errors, which heats up
significantly the regions with temperature $\leq 10^2$K to $T\sim 10^2-10^4$%
K in the WENO-E calculation. In contrast, since the entropy
equation is capable of tracking the temperature field with high
accuracy and hence the spurious heating is minimized, the cold
unshocked regions occupy more volumes in the WENO-S calculation.

Fig. 10 gives an example of density, velocity and temperature
distributions along a randomly chosen lines of sight. To
demonstrate further the difference between the WENO-E and WENO-S
codes, we plot the results from both calculations in each panel.
Once again, the numerical heating in the pre-shock or unshocked
regions is clearly seen in this plot. Our result is compatible
with that of Ryu et al. (1993) for a purely baryonic universe.
Moreover, it is also noted from Fig. 10 that there are not
significant differences between the density and velocity fields in
the WENO-S and WENO-E calculations, while for the temperature
distribution, the difference occurs in the regions of cold gas,
$T\le 10^4$K. This demonstrates that the internal energy
corrections made in cold regions (mostly in unshocked gas) by the
modified entropy equation have little dynamical effect on the flow
structure except for the internal energy or temperature fields.

In Fig. 11, we present an alternative view of the difference
between the WENO-E and WENO-S codes by plotting histograms of the
volume-weighted (upper panel) and mass-weighted (lower panel)
temperature on the $192^3$ grids. Clearly, the artificial
numerical heating is serious in the regions occupied by the
low-temperature gas in the WENO-E calculation, however it is less
significant in the fraction of mass. Another illustration
indicating such a difference can be given by contour plots of the
volume fraction with given temperature and density. As displayed
in Fig. 12, the volume difference only occurs in the
low-density/low temperature regions with $\rho/\bar{\rho}\le 1$
and $T\le 10^4$K, typically around $\rho/\bar{\rho}\sim 0.1$,
$T\sim 10^3$K. It should be noted that, although the spurious
heating does exist also in the standard WENO code, it is weaker
than those in low-order schemes, e.g., the second order TVD code
of Ryu et al. (1993). This benefit clearly comes from the higher
order accuracy acquired in our fifth-order WENO scheme, which
leads to smaller numerical errors in the calculations. However,
for realistic cosmological simulations, such a difference may not
be significant while the radiation field is taken into account.
For instance, in the presence of ionizing UV background, the
baryonic intergalactic medium could be heated up to $\sim 10^4$K.
The simulation including the radiative heating-cooling and
ionization and relevant statistical analysis will be presented
elsewhere.

\section{Concluding Remarks}

In this paper, we have described a newly developed hybrid
cosmological hydrodynamic code based on weighted essentially
non-oscillatory (WENO) schemes for the Euler system of
conservation laws. We implement the fifth order finite difference
WENO to solve the inviscid fluid dynamics on a uniform Eulerian
grid combining with a third order low storage Runge-Kutta TVD
scheme for advancing in time. In order to solve the cosmological
problem involving both collisional baryonic matter and
collisionless dark matter, we incorporate the particle-mesh method
for computing the self-gravity into our cosmological code.

The code has been subjected to a number of tests for its accuracy
and convergence. As expected, the WENO scheme demonstrates its
capacity of capturing shocks and producing sharp and
non-oscillatory discontinuity transition without generating
oscillations. In comparison with other existing hydrodynamic codes
such as the TVD or PPM schemes, one striking feature of the WENO
code is that it retains higher order accuracy in smooth regions
including at smooth extrema even in multidimensions, and yet it is
still highly stable and robust for strong shocks. In performance,
the WENO scheme needs more floating point operations per cell than
those of the PPM and TVD schemes. However, in compensating for
twice or more loss of the computational speed, the WENO scheme
achieves both higher order accuracy and convergence rate than PPM
and TVD codes according to our numerical tests.

In the presence of gravity, the hydrodynamics become more
challenging than that without gravity due to the highly
non-linearity of gravitational clustering. One serious problem
encountered in many cosmological applications is the so called
high Mach number problem. To address this problem, we have
incorporated an extra technique into our cosmological WENO/PM
code, which is actually a combination of the dual energy algorithm
(Bryan et al. 1995) and the energy-entropy algorithm (Ryu et al.
1993). Namely, instead of solving the internal energy equation in
regions free of shocks as was done in the dual energy algorithm
(Bryan et al. 1995), we solve the modified entropy equation (Ryu
et al. 1993), which takes a conservative form and can be easily
solved using the standard WENO scheme. This improvement over our
hydrodynamic WENO code ensures an accurate tracking of the
temperature field in regions free of shocks.

It is pointed out that the high order WENO discretization, e.g.,
the fifth order WENO scheme adopted in this paper, introduces a
quite small numerical viscosity, which does not lead to a
significant violation of energy conservation in the presence of
gravitational fields. While for a second-order TVD scheme, the
numerical diffusion is no longer negligible. In order to have a
better conservation of the total energy, it is usually corrected
by adding a compensation term in the gravitational force term (Ryu
et al, 1993).

The Euler hydrodynamics on fixed meshes have several distinct
advantages which includes simplicity for implementation, easy data
parallelization, relatively low floating point cost, large dynamic
range in mass and high resolution of shock capturing. In
particular, the WENO scheme can also achieve a higher accuracy on
coarse meshes and a better resolution on the largest meshes
allowed by available memory. To suit for large simulations of the
cosmological problem, further improvement of the hydrodynamic WENO
code is needed in its implementation on distributed memory
computers. The parallel version of the code based on the
Message-Passing Interface (MPI) has been under development.

\acknowledgments

LLF acknowledges support from the National Natural Science
Foundation of China (NNSFC) and National Key Basic Research
Science Foundation. CWS and MZ acknowledge the support by NNSFC
grant 10028103 while CWS is in residence at the Department of
Mathematics, University of Science and Technology of China, Hefei,
Anhui 230026, P.R. China. Additional support for CWS is provided
by ARO grant DAAD19-00-1-0405 and NSF grant DMS-0207451.

\appendix

\section{Low Storage Runge-Kutta Scheme}

General low-storage Runge-Kutta schemes can be written in the form
\begin{eqnarray}  \label{s2}
du^{(i)} & = & A_i du^{(i-1)}+ \Delta t L(u^{(i-1)}, t^n +
\alpha_i \Delta t)
\nonumber \\
u^{(i)} & = & u^{(i-1)} + B_{i} du^{(i)}, \qquad i=1, ..., m \\
u^{(0)} &= &u^n, \,\,\, u^{(m)} = u^{n+1}, \,\,\, A_1 =0 .
\nonumber
\end{eqnarray}
Only $u$ and $du$ must be stored, resulting in two storage units
for each variable, instead of three storage units for equation
(\ref{s1}). The third order nonlinearly stable version we use,
Gottlieb \& Shu (1998), has $m=3$ in (\ref{s2}) with
\begin{eqnarray}  \label{s3}
z_1 & = & \sqrt{36 c_2^4+36 c_2^3-135 c_2^2+84 c_2-12}  \nonumber \\
z_2 & = & 2 c_2^2+c_2-2  \nonumber \\
z_3 & = & 12 c_2^4-18 c_2^3+18 c_2^2-11 c_2+2  \nonumber \\
z_4 & = & 36 c_2^4-36 c_2^3+13 c_2^2-8 c_2+4  \nonumber \\
z_5 & = & 69 c_2^3-62 c_2^2+28 c_2-8  \nonumber \\
z_6 & = & 34 c_2^4-46 c_2^3+34 c_2^2-13 c_2+2  \nonumber \\
B_1 & = & c_2 \\
B_2 & = & \frac{12 c_2 (c_2-1) (3 z_2-z_1)-(3 z_2-z_1)^2}{ 144 c_2
(3 c_2-2)
(c_2-1)^2}  \nonumber \\
B_3 & = & \frac{-24 (3 c_2-2) (c_2-1)^2}{ (3 z_2-z_1)^2-12 c_2
(c_2-1) (3
z_2-z_1)}  \nonumber \\
A_2 & = & \frac{-z_1 (6 c_2^2-4 c_2+1)+3 z_3}{ (2 c_2+1) z_1-3
(c_2+2)(2c_2-1)^2}  \nonumber \\
A_3 & = & \frac{-z_4 z_1+108 (2 c_2-1) c_2^5-3 (2 c_2-1) z_5}{ 24
z_1 c_2
(c_2-1)^4+72 c_2 z_6+72 c_2^6 (2 c_2-13)}  \nonumber \\
\alpha_1 & = & 0  \nonumber \\
\alpha_2 & = & B_1  \nonumber \\
\alpha_3 & = & B_1 + B_2 (A_2 +1)  \nonumber
\end{eqnarray}
where $c_2 = 0.924574$.

\clearpage

\begin{figure}
\figurenum{1} \epsscale{0.9}
\plotfiddle{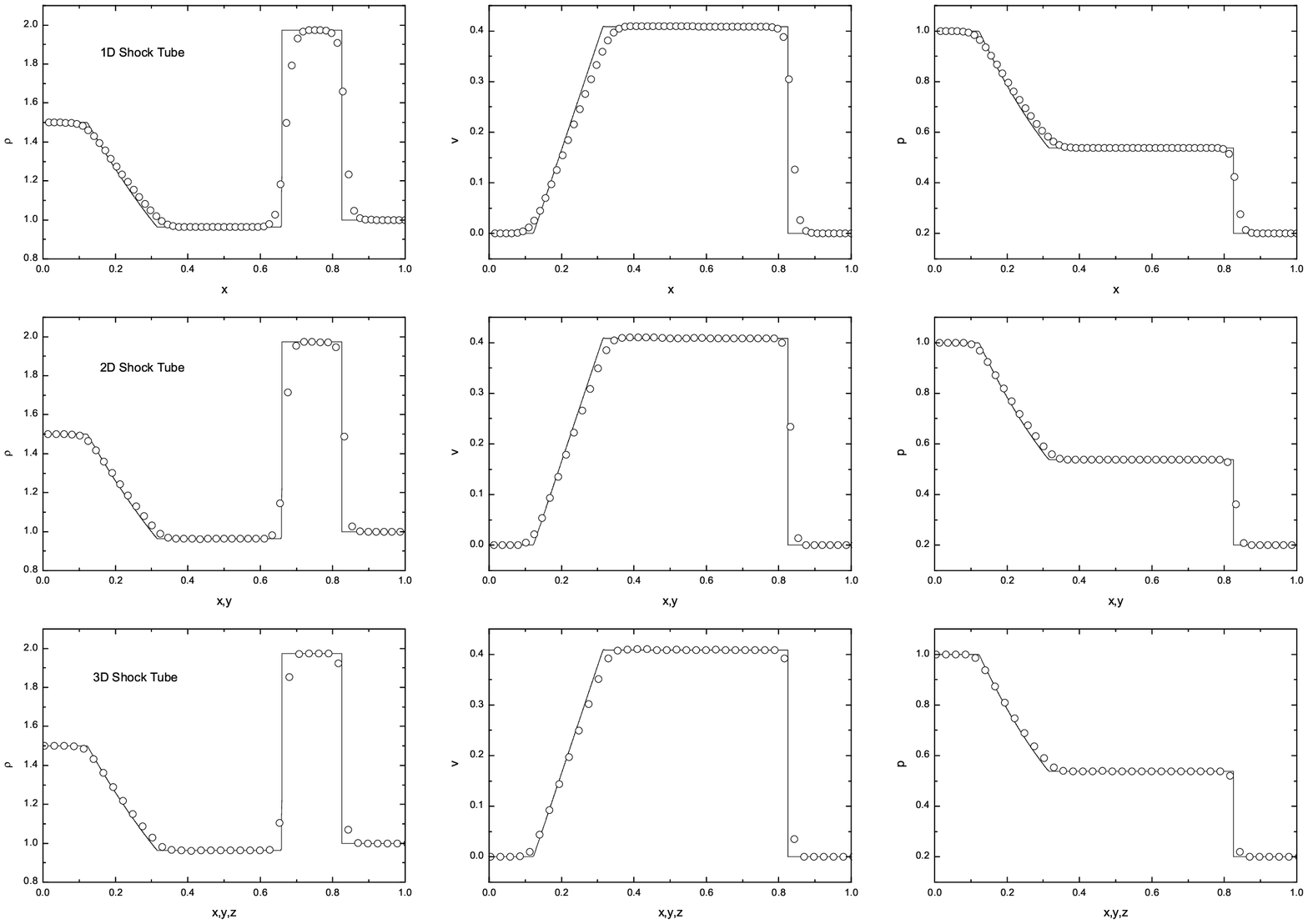}{20.cm}{90}{90}{90}{250.}{-40.} \caption[]
{Density (left), velocity (center) and pressure (right) for the
Sod shock tube tests in one-, two- and three-dimensions. Open
circles are given by the numerical solution output at $t=0.195$
using the fifth order WENO scheme, solid lines represent the
analytical solutions.}
\end{figure}

\begin{figure}
\figurenum{2} \epsscale{0.9}
\plotfiddle{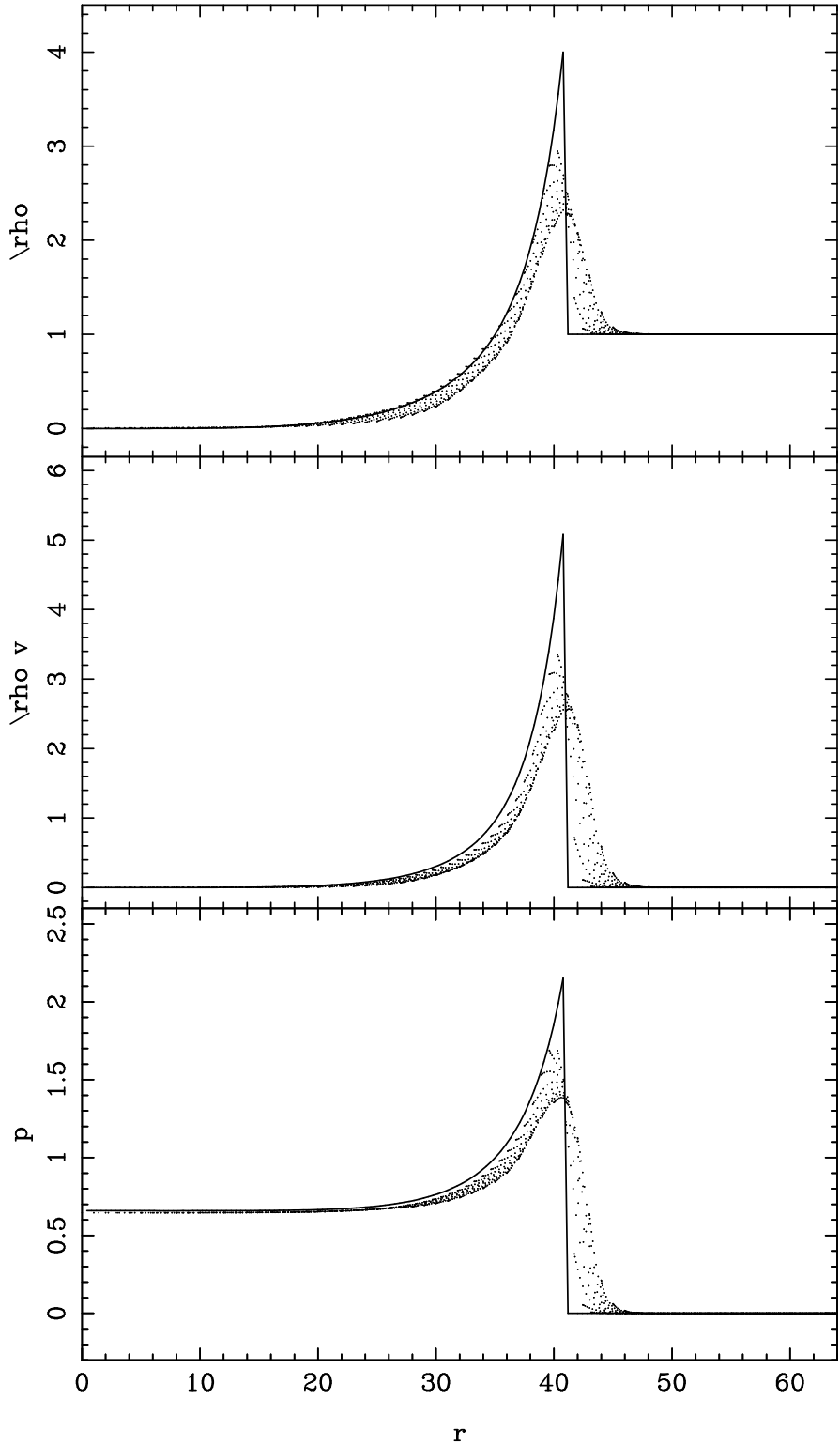}{20.cm}{0}{90}{90}{-280.}{-40.}
\caption[]{The density (top), momentum density (middle) and
pressure (bottom) in the three-dimensional spherical Sedov blast
wave test at t=9.22. The computation was performed on a $256^3$
grid, and the scattered points are plotted by the projection of
the results on the Cartesian grids onto the spherical radial
coordinates. The solid lines represent the analytical solutions.}
\end{figure}

\begin{figure}
\figurenum{3}\epsscale{1.0}
\plotfiddle{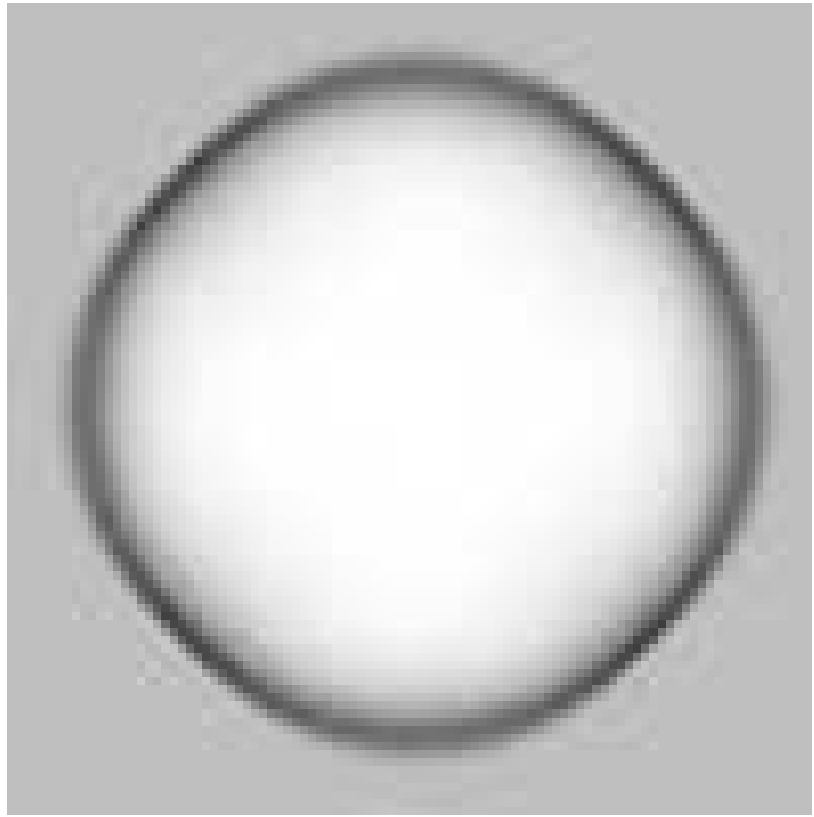}{20.cm}{0}{90}{90}{-280.}{-50.}
\caption[]{The gray image of density distribution in a slice
across the explosion point in the three-dimensional spherical
Sedov blast wave at t=9.22.} \end{figure}

\begin{figure}
\figurenum{4}\epsscale{0.8}
\plotfiddle{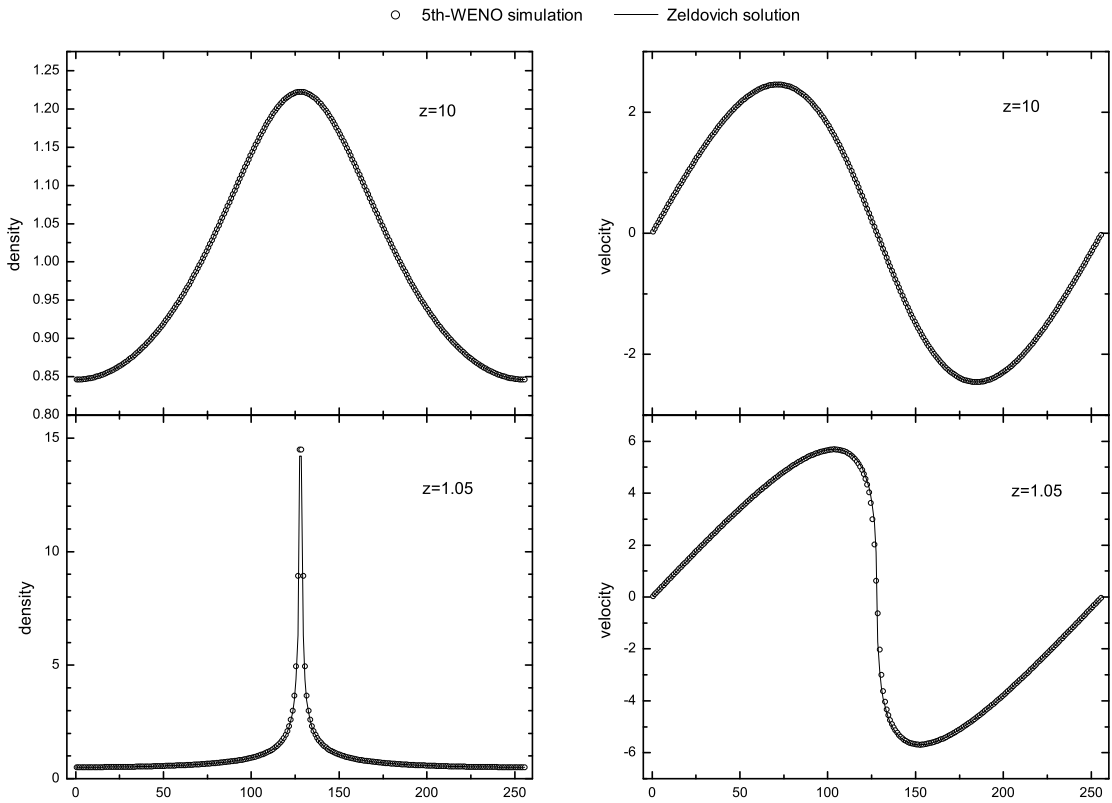}{20.cm}{0}{80}{80}{-270.}{30.} \caption[]
{The one-dimensional  Zeldovich Pancake test at redshifts z=10 and
z=1.05 prior to the caustics formation. The results from a 256
grids run (open circle) are plotted against the exact solutions
(solid line).}
\end{figure}

\begin{figure}
\figurenum{5}\epsscale{0.9}
\plotfiddle{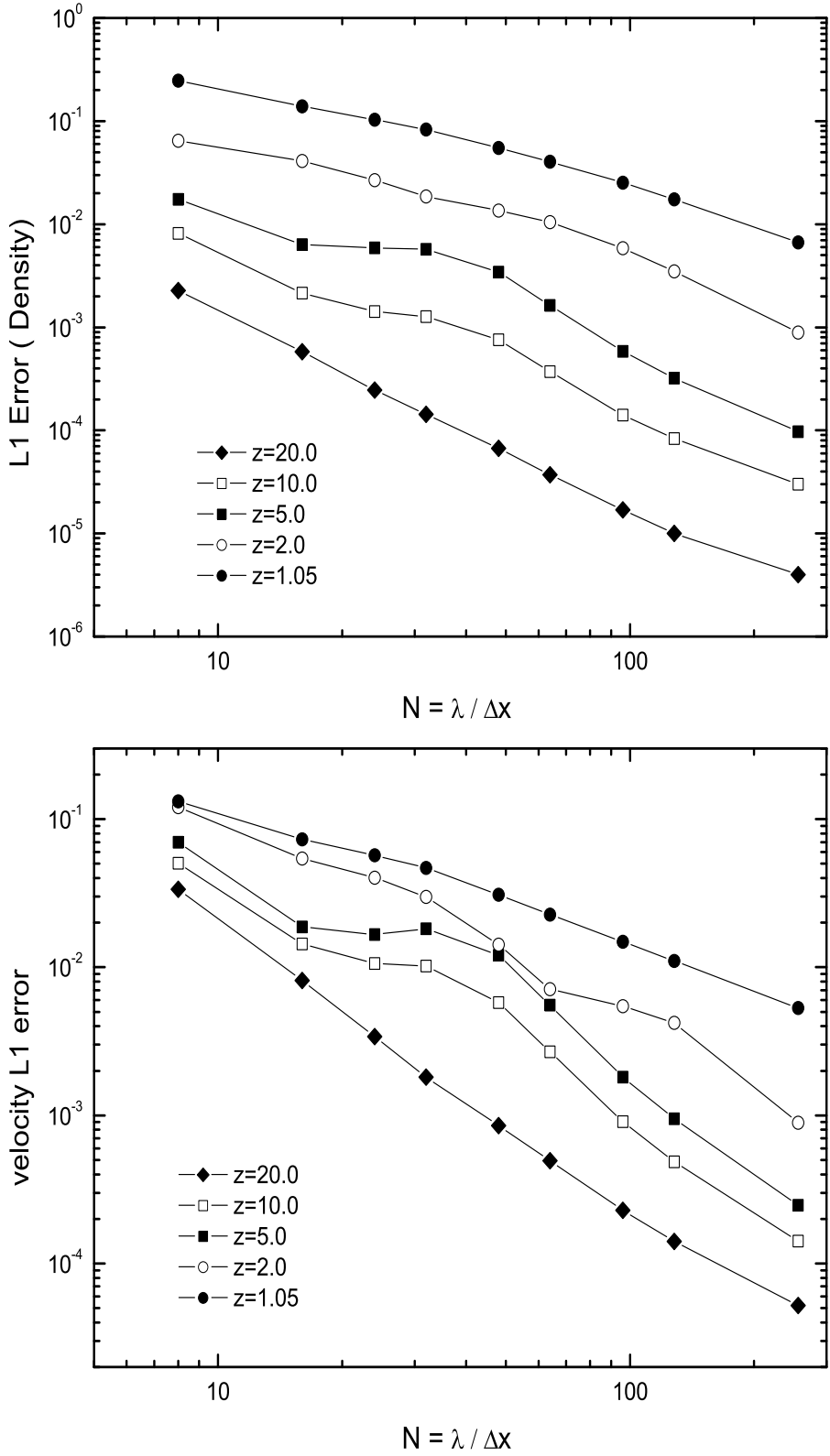}{20.cm}{0}{90}{90}{-240.}{-60.} \caption[]
{The $L^1$ error norm of one-dimensional Zeldovich pancake tests
varying with the number of zones $N$ for a fixed perturbation
wavelength. The results for density (upper panel) and velocity
(lower panel) are plotted at different redshifts.}
\end{figure}

\begin{figure}
\figurenum{6}\epsscale{0.9}
\plotfiddle{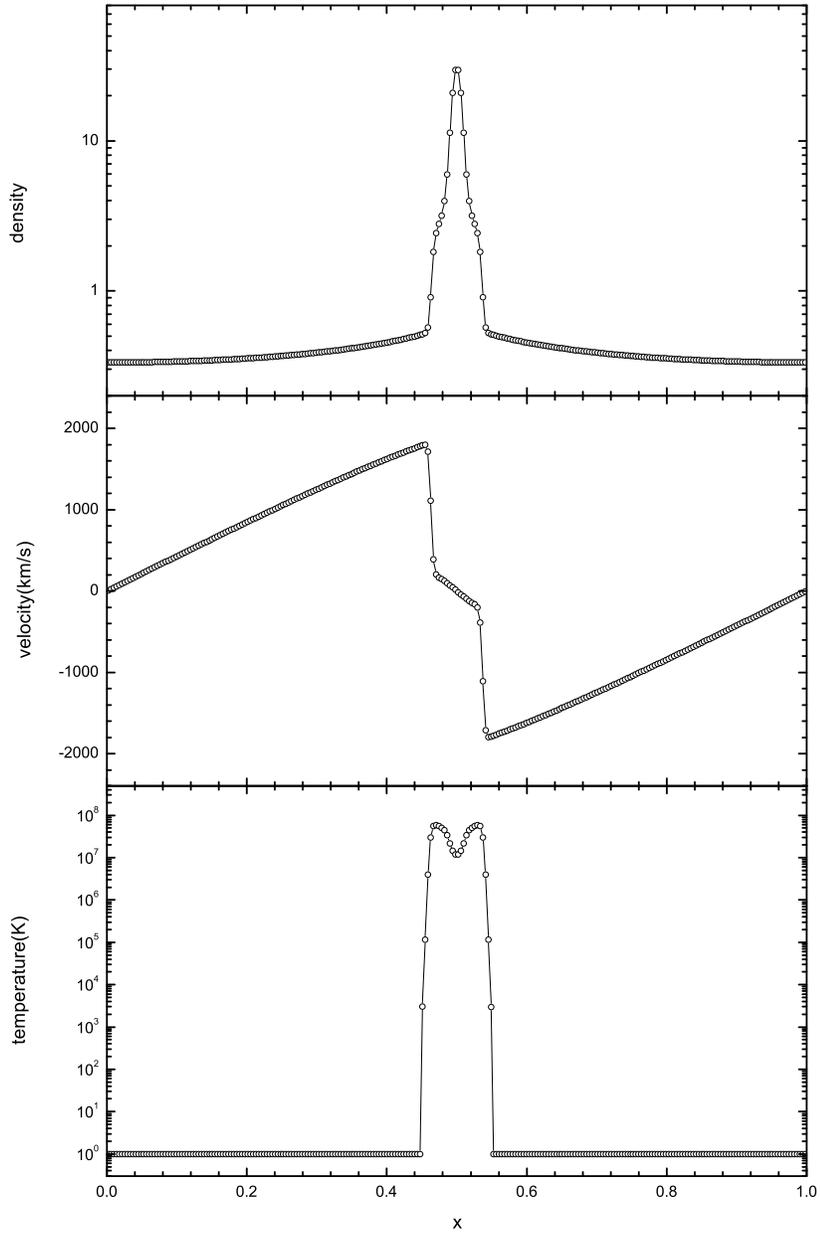}{20.cm}{0}{90}{90}{-190.}{-10.}
\caption[]{The density (top), velocity (middle) and temperature
(bottom) for a one-dimensional Zeldovich pancake at redshift z=0.
The results are drawn from a 256 grid run.}
\end{figure}

\begin{figure}
\figurenum{7}\epsscale{1.0}
\plotfiddle{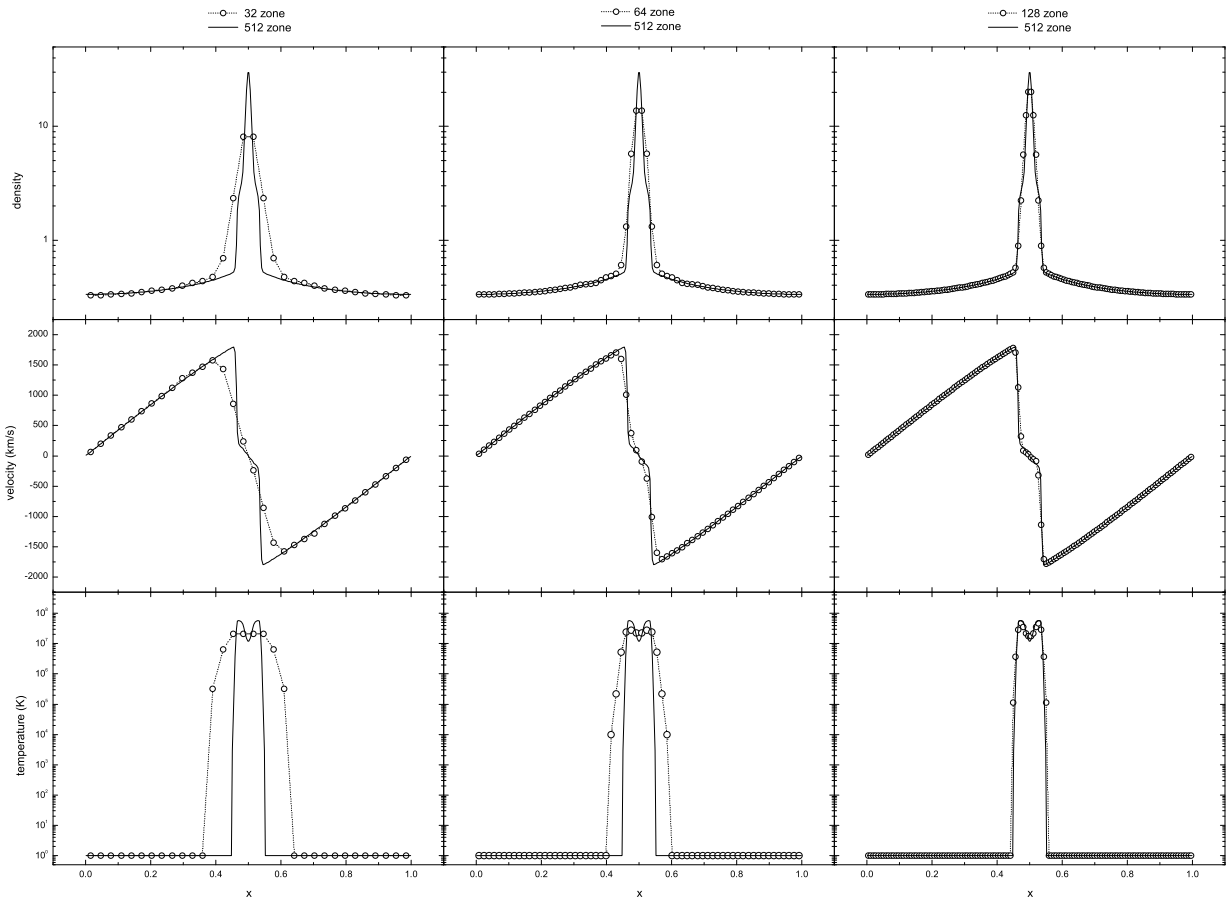}{20.cm}{0}{90}{90}{-300.}{-10.}
\caption[]{The density (top), velocity (middle) and temperature
(bottom) for a one-dimensional Zeldovich pancake at redshift z=0.
Open circles with dotted lines are obtained from the runs with 32,
64, 128 grids.  The solid lines are for the solution computed with
a 512 grid run.} \end{figure}

\begin{figure}
\figurenum{8}\epsscale{0.9}
\plotfiddle{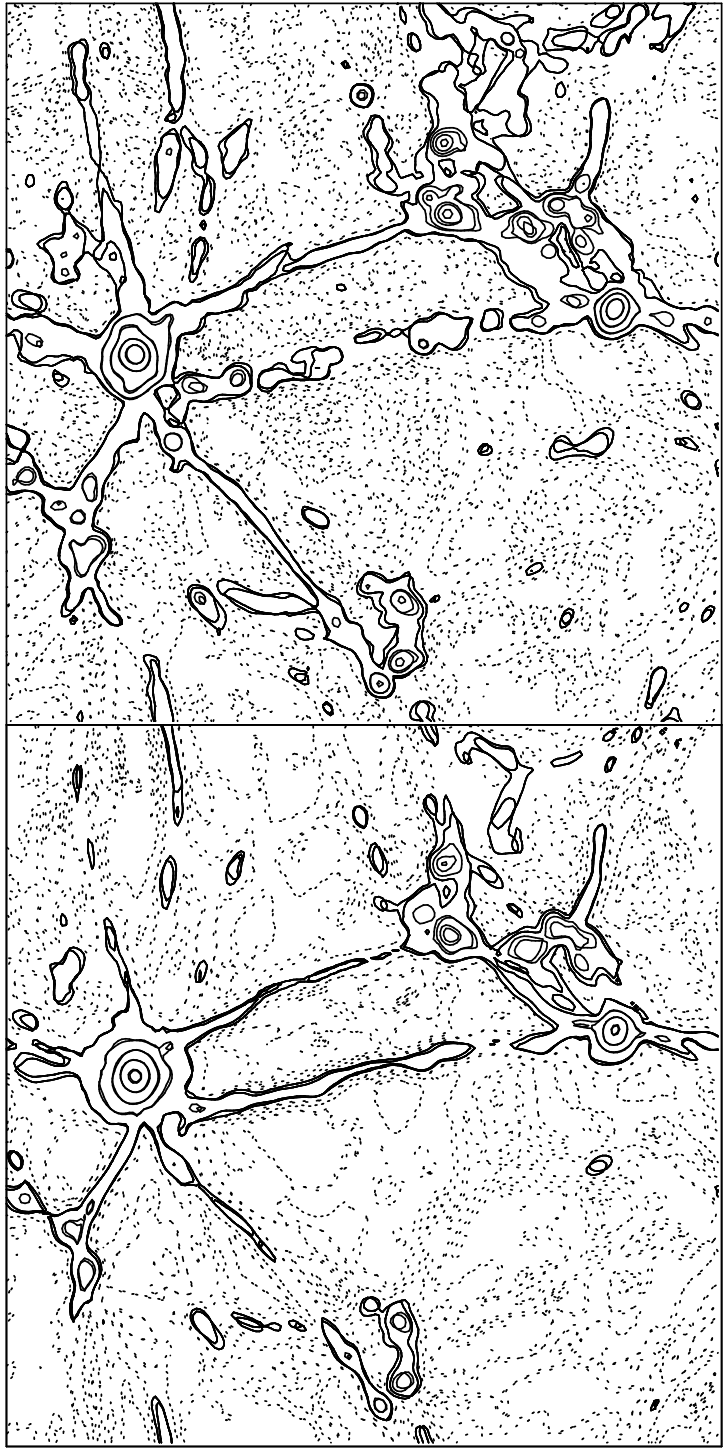}{20.cm}{0}{90}{90}{-300.}{-60.}
\caption[]{Gas density (lower panel) and CDM density (upper panel)
contour plots for a slice of 0.26h$^{-1}$Mpc thickness in a
192$^3$ grid at $z=1.5$. The solid line contours represent
overdense regions with $\rho/\bar{\rho}\geq 1$, and the dotted
lines represent underdense regions with $\rho/\bar{\rho}\leq 1$.}
\end{figure}

\begin{figure}
\figurenum{9}\epsscale{0.9}
\plotfiddle{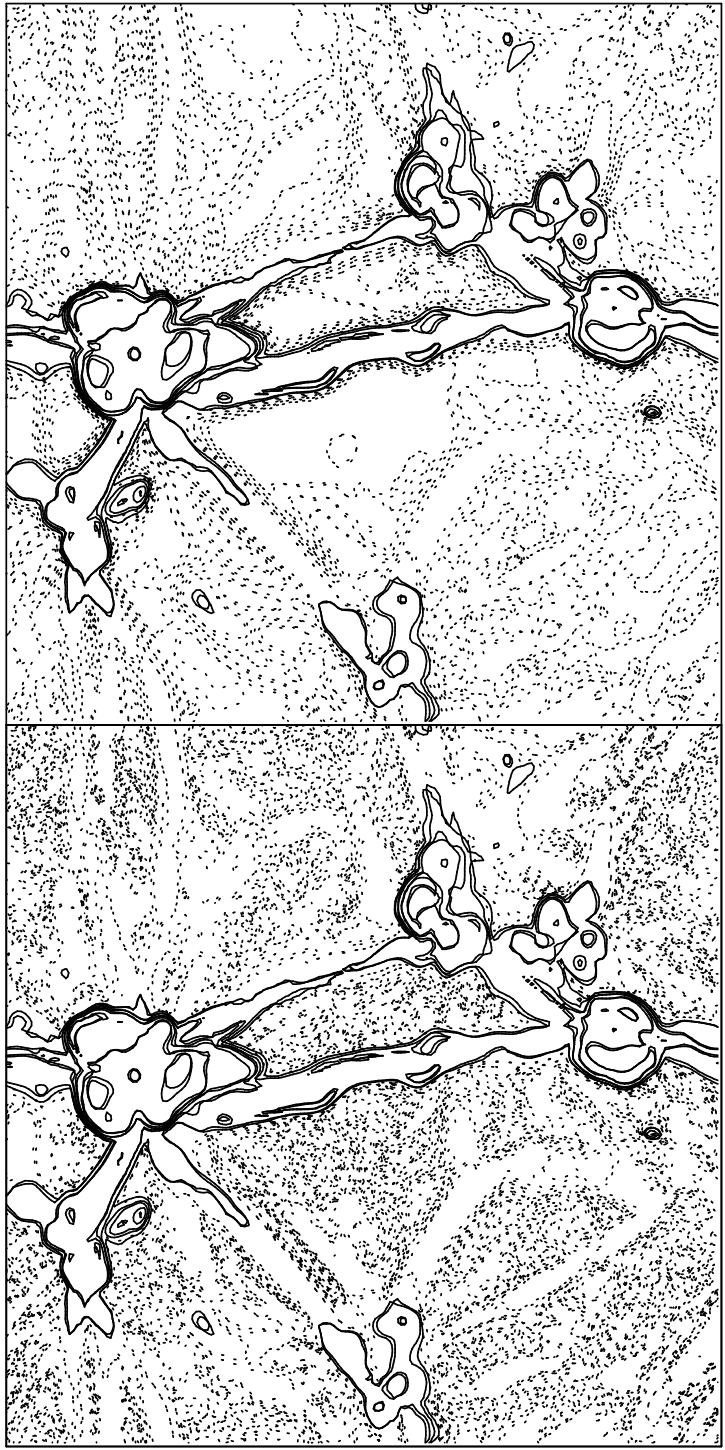}{20.cm}{0}{90}{90}{-300.}{-60.}
\caption[]{Comparison of temperature contours for the WENO-E
(lower panel) and WENO-S (upper panel) for a slice of
0.26h$^{-1}$Mpc thickness in a 192$^3$ grid at $z=1.5$. The solid
line contours represent hot regions with $T\ge 10^4$K and the
dotted lines represent cold region with $T<10^4$K.}
\end{figure}

\begin{figure}
\figurenum{10}\epsscale{0.9}
\plotfiddle{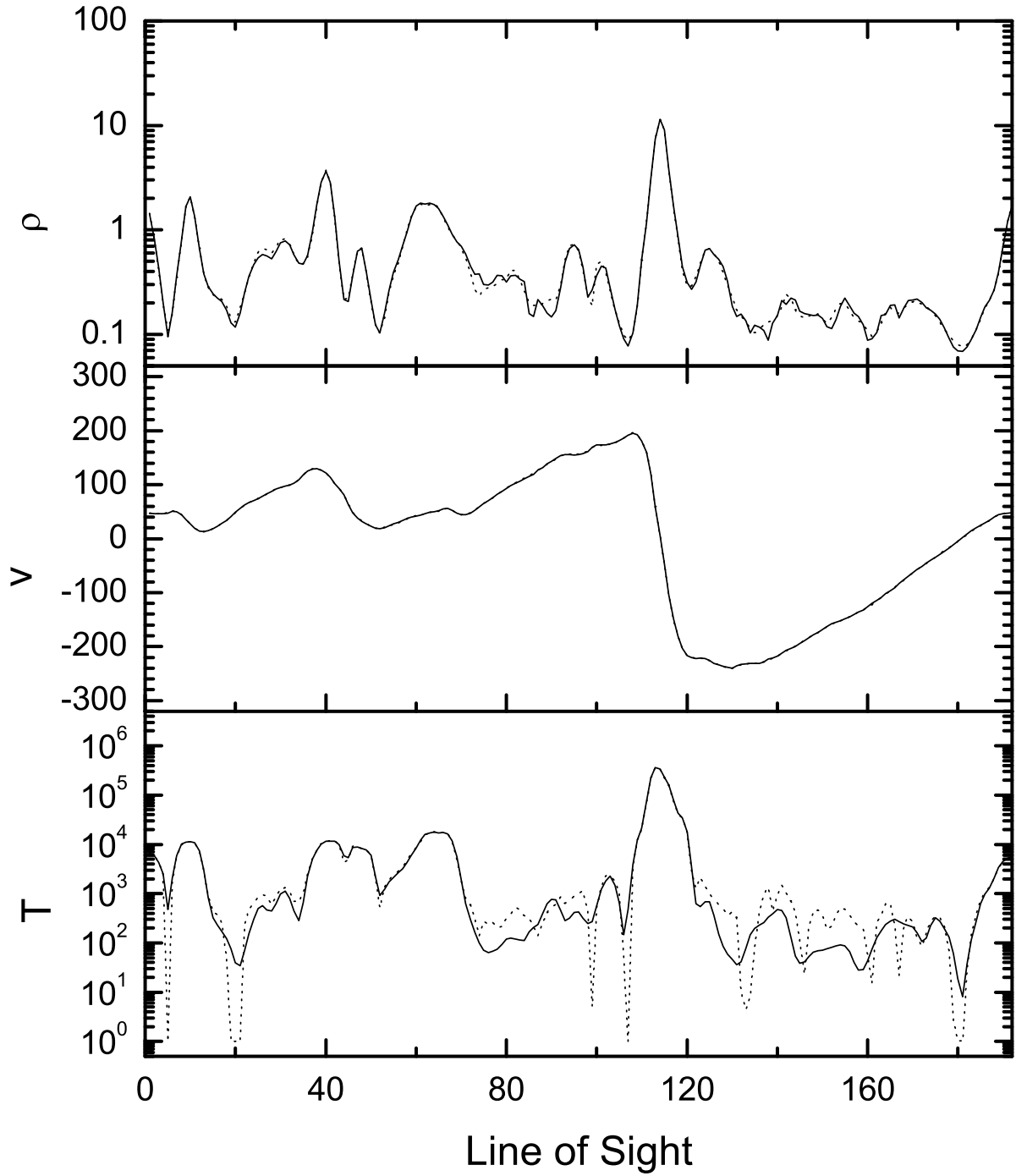}{20.cm}{0}{90}{90}{-270.}{-10.}
\figcaption{Density, velocity and temperature distributions along
a randomly chosen lines of sight at $z=1.5$. The solid lines are
for the WENO-S simulation and the dotted lines for the WENO-E
simulation.} \end{figure}

\begin{figure}
\figurenum{11}\epsscale{0.9}
\plotfiddle{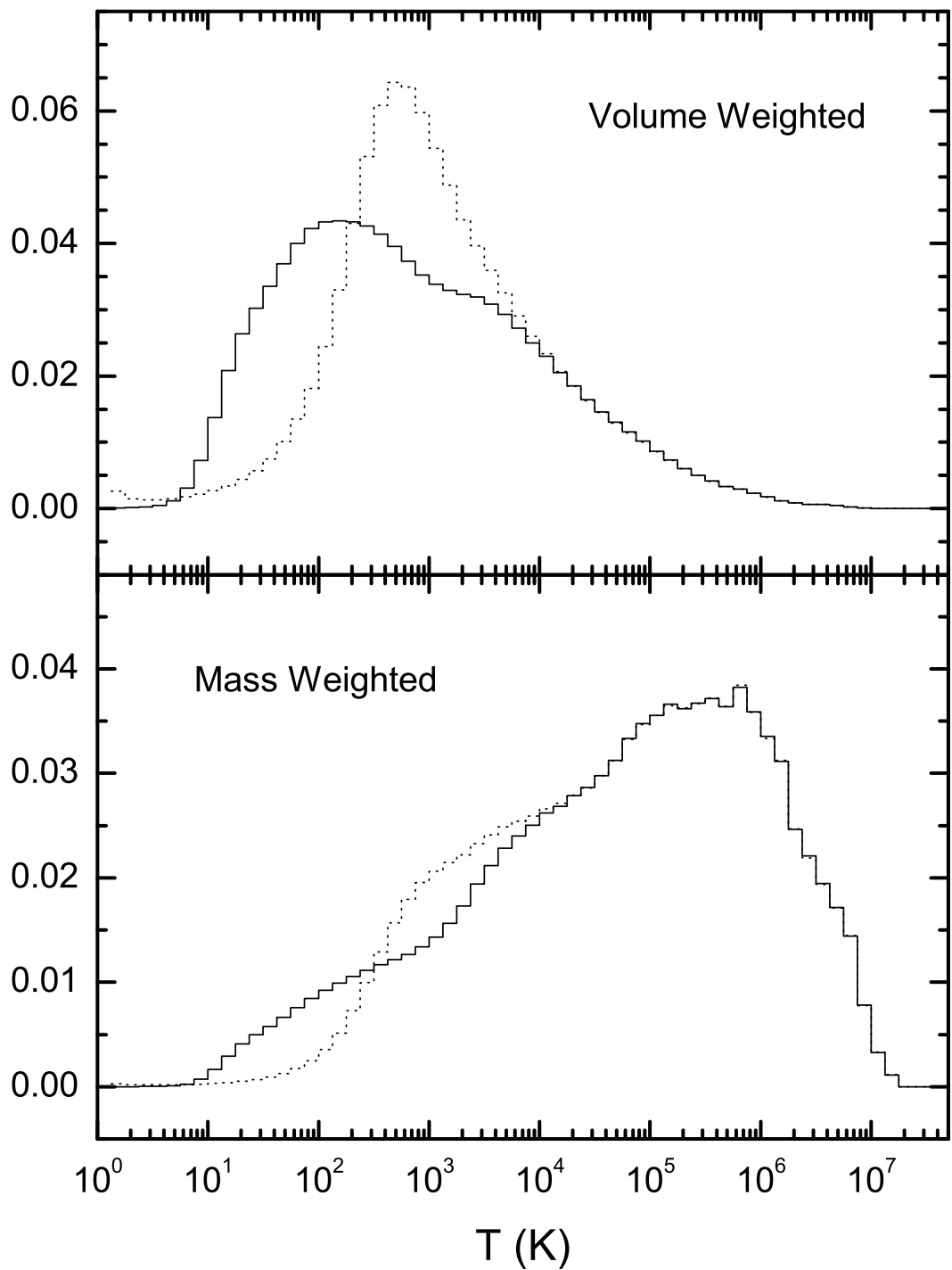}{20.cm}{0}{90}{90}{-270.}{-10.}
\caption[]{The cell temperature histogram in a 192$^3$ grid at
$z=1.5$. The top panel is for the volume-weighted cell temperature
and the lower panel for the mass-weighted cell temperature. In
both panels, the solid lines are given by the WENO-S simulation
and the dotted lines given by the WENO-E.}\end{figure}

\begin{figure}
\figurenum{12}\epsscale{1.0}
\plotfiddle{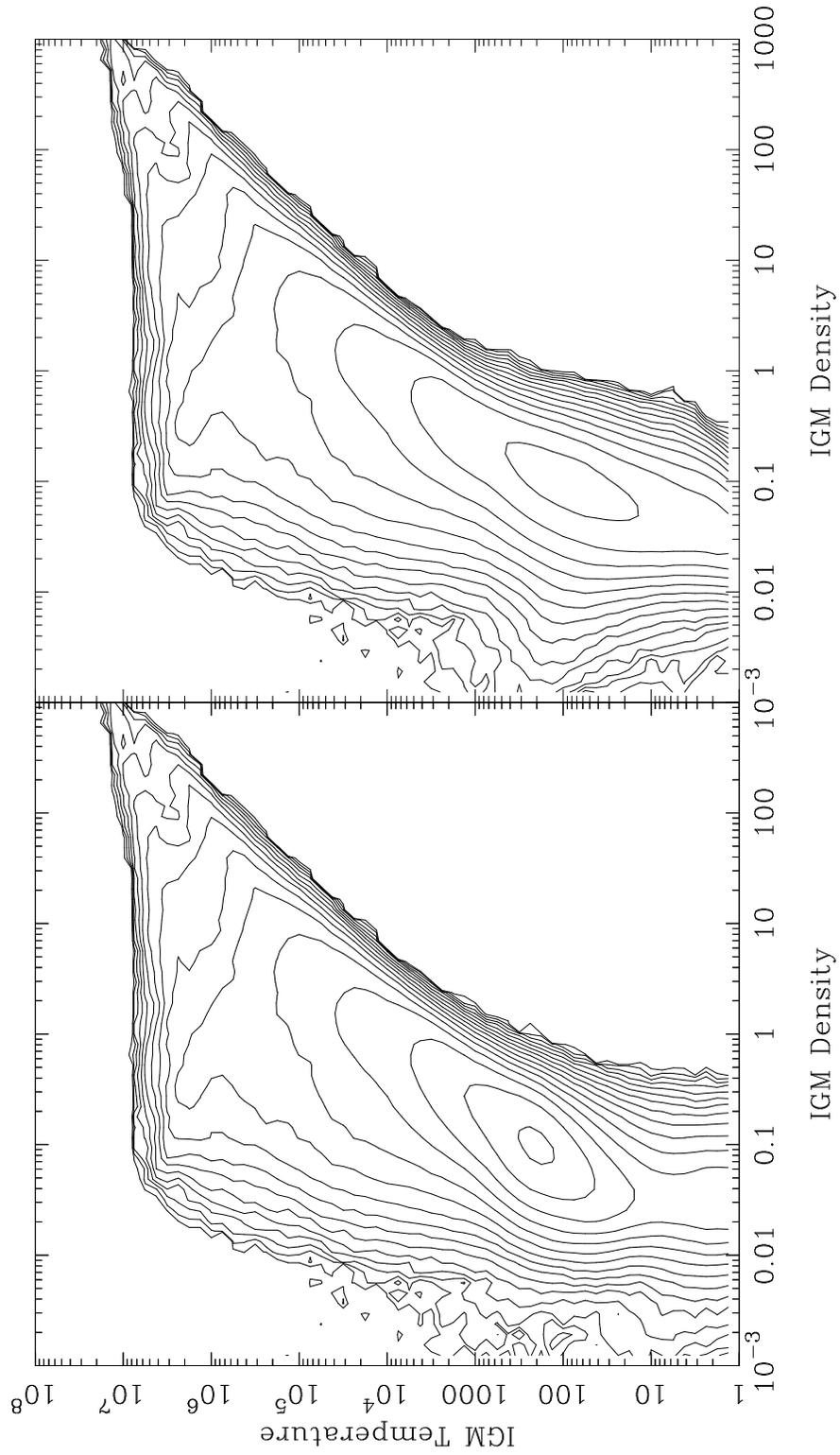}{20.cm}{0}{100}{100}{-300.}{-100.}
\caption[]{Contour plot of the volume with given temperature and
density at $z=1.5$. The left panel is for the calculation with the
WENO-E and the right panel with the WENO-S. }
\end{figure}

\end{document}